\documentclass[aps,prl,reprint,superscriptaddress,longbibliography]{revtex4-1}

\usepackage{graphicx}
\usepackage{dcolumn}
\usepackage{bm}
\usepackage{amsfonts}
\usepackage{amssymb}
\usepackage{amsmath}
\usepackage{graphicx}
\usepackage{color}
\usepackage{array}
\usepackage{dcolumn} 
\usepackage{bm} 
\usepackage{amsthm}
\usepackage{latexsym}
\usepackage{float}
\usepackage{tabularx}

\begin{document}


\title{Dark breathers in a normal dispersion optical microresonator}


\author{Chengying Bao}
\email{bao33@purdue.edu}
\affiliation{School of Electrical and Computer Engineering, Purdue University, 465 Northwestern Avenue, West Lafayette, Indiana 47907-2035, USA}

\author{Yi Xuan}
\affiliation{School of Electrical and Computer Engineering, Purdue University, 465 Northwestern Avenue, West Lafayette, Indiana 47907-2035, USA}
\affiliation{Birck Nanotechnology Center, Purdue University, 1205 West State Street, West Lafayette, Indiana 47907, USA}

\author{Daniel E. Leaird}
\affiliation{School of Electrical and Computer Engineering, Purdue University, 465 Northwestern Avenue, West Lafayette, Indiana 47907-2035, USA}

\author{Minghao Qi}
\affiliation{School of Electrical and Computer Engineering, Purdue University, 465 Northwestern Avenue, West Lafayette, Indiana 47907-2035, USA}
\affiliation{Birck Nanotechnology Center, Purdue University, 1205 West State Street, West Lafayette, Indiana 47907, USA}

\author{Andrew M. Weiner}
\email{amw@purdue.edu}
\affiliation{School of Electrical and Computer Engineering, Purdue University, 465 Northwestern Avenue, West Lafayette, Indiana 47907-2035, USA}
\affiliation{Birck Nanotechnology Center, Purdue University, 1205 West State Street, West Lafayette, Indiana 47907, USA}
\begin{abstract}
Breathers are localized waves, that are periodic in time or space. The concept of breathers is useful for describing many physical systems including granular lattices, Bose-Einstein condensation, hydrodynamics, plasmas and optics. Breathers could exist in both the anomalous and the normal dispersion regime. However, the demonstration of optical breathers in the normal dispersion regime remains elusive to our knowledge. Kerr comb generation in optical microresonators provides an array of oscillators that are highly coupled via the Kerr effect, which can be exploited to explore the breather dynamics. Here, we present, experimentally and numerically, the observation of dark breathers in a normal dispersion silicon nitride microresonator. By controlling the pump wavelength and power, we can generate the dark breather, which exhibits an energy exchange between the central lines and the lines at the wing. The dark breather breathes gently and retains a dark-pulse waveform.  A transition to a chaotic breather state is also observed by increasing the pump power. These dark breather dynamics are well reproduced by numerical simulations based on the Lugiato-Lefever equation. The results also reveal the importance of dissipation to dark breather dynamics and give important insights into instabilities related to high power dark pulse Kerr combs from normal dispersion microreosnators.
\end{abstract}


\maketitle
The interaction between nonlinearity and dispersion leads to various localized solutions to the nonlinear Schr$\ddot{o}$dinger equation (NLSE). Solitons are one of the most well-known solution \cite{Zakharov_JETP1972} and have advanced various research arenas such as Bose-Einstein condensates (BEC) \cite{Phillips_Science2000} and optics \cite{Dudley_NP2010Ten}. Notably, soliton mode-locking has enabled the coherent Kerr comb generation from optical microresonators recently \cite{Kippenberg_Nature2007,Kippenberg_Science2011,
Kippenberg_NP2014temporal} (a Kerr comb is an array of equally spaced spectral lines generated by cascaded four-wave-mixing in high-$Q$ microresonators, see ref. \cite{Kippenberg_Science2011} for a review). Breathers are another type of solution to the NLSE that occur widely in physical systems \cite{Kuznetsov_PR1986soliton,Ma1979,Akhmediev1986,Dudley_NP2010,Smerzi_PRL2001}. Different from solitons which preserve their shape as they propagate, breathers feature periodic variation in time or space. Systems in which breathers may exist include BEC \cite{Smerzi_PRL2001}, granular lattices \cite{Willis1998}, hydrodynamics \cite{Akhmediev_PRL2011Rogue}, optics \cite{Dudley_NP2010,Dudley_NP2014}. Breathers are attracting growing attention recently due to their close relationship to the growth and decay of the extreme events known as rogue waves \cite{Jalali_Natue2007,Dudley_NP2010,Akhmediev_PRL2011Rogue,Akhmediev_PRX2012,Dudley_NP2014,Dudley_NC2016}.

Optical systems are widely used to study breathers, due to the ease to tailor the linear and nonlinear properties of the systems. Optical breathers were first studied in conservative systems, namely optical fibres \cite{Dudley_NP2010,Akhmediev_PRX2012,Dudley_NP2014,Dudley_NC2016}. Recently optical breathers have also been observed in dissipative systems such as externally pumped fiber cavities \cite{Coen_OE2013} and microresonators \cite{Weiner_PRL2016,Gaeta_NC2017,Kippenberg_arXiv2016Breather,Kippenberg_arXiv2017inter}. Kerr comb generation in optical microresonators \cite{Kippenberg_Nature2007,Kippenberg_Science2011} constitutes an excellent experimental platform to explore breather dynamics, as microresonators provide essentially infinite propagation distance where breathers are densely sampled every round-trip. Furthermore, the wide frequency spacing for Kerr combs generated from microresonator \cite{Kippenberg_Science2011} makes it easy to separate individual lines and study their breathing dynamics \cite{Weiner_PRL2016}. To date, optical breathers have been mainly explored in the anomalous dispersion regime (for which nonlinearity and dispersion have the same sign). Nonlinear dynamics in the normal dispersion regime (opposite sign for nonlinearity and dispersion) are also important; for instance they help to improve pulse energy of mode-locked lasers \cite{Wise_LPR2008}. Dark solitons, which have been reported in optics \cite{Weiner_PRL1988}, BEC \cite{Lewenstein_PRL1999} and hydrodynamics \cite{Akhmediev_PRL2013} may exist in this regime. However, experimental observation of optical breathers in the normal dispersion regime remains elusive, to our knowledge. Discrete dark breathers have been shown to exist in granular lattices (e.g., Fermi-Pasta-Ulam lattice and Klein-Gordon lattice) \cite{Daraio_PRE2014,Romero_NJP2002}. Nevertheless, many questions about dark breathers, such as their dynamics in highly driven cases, remain open. Kerr comb generation in normal dispersion microresonators \cite{Weiner_NP2015mode,Weiner_Optica2014,Matsko_OL2014,Gorodetsky_OE2015,Wong_PRL2015,
Gaeta_OE2016dynamics,Gelens_PRA2016,Chembo_PRA2014} can provide a unique opportunity to study optical dark breathers. Numerical simulations \cite{Chembo_PRA2014,Gelens_PRA2016,Weiner_NP2015mode} and preliminary experiments \cite{Weiner_NP2015mode} suggest that optical dark breathers may exist in normal dispersion microresonators. In   this Article, we present the first clear observation and detailed characterization of optical dark breathers in the normal dispersion regime, which we obtain by destabilizing a dark pulse Kerr comb in a silicon-nitride (SiN) microresonator. The dark breather breathes relatively gently and features an energy exchange between comb lines. Furthermore, the breathing behaviour changes abruptly from line to line. The observed dark breather dynamics in dissipative cavities is governed by the Lugiato-Lefever equation (LLE), i.e., driven-damped NLSE \cite{Lugiato_PRL1987spatial,Coen_OL2013modeling} and differs substantially from the breathing ``double periodic" solution in the NLSE with normal dispersion \cite{Akhmediev_PRA1993}. In the highly driven regime, we observe that the dark breather in our dissipative cavity undergoes a period-tripling bifurcation and then a transition into a chaotic state. This behavior is prohibited in systems governed by the NLSE. The experimental observation of dark breathers is confirmed by rigorous comparison with numerical simulations.

\begin{figure*}[t]
\centering
\includegraphics[width=1.7\columnwidth]{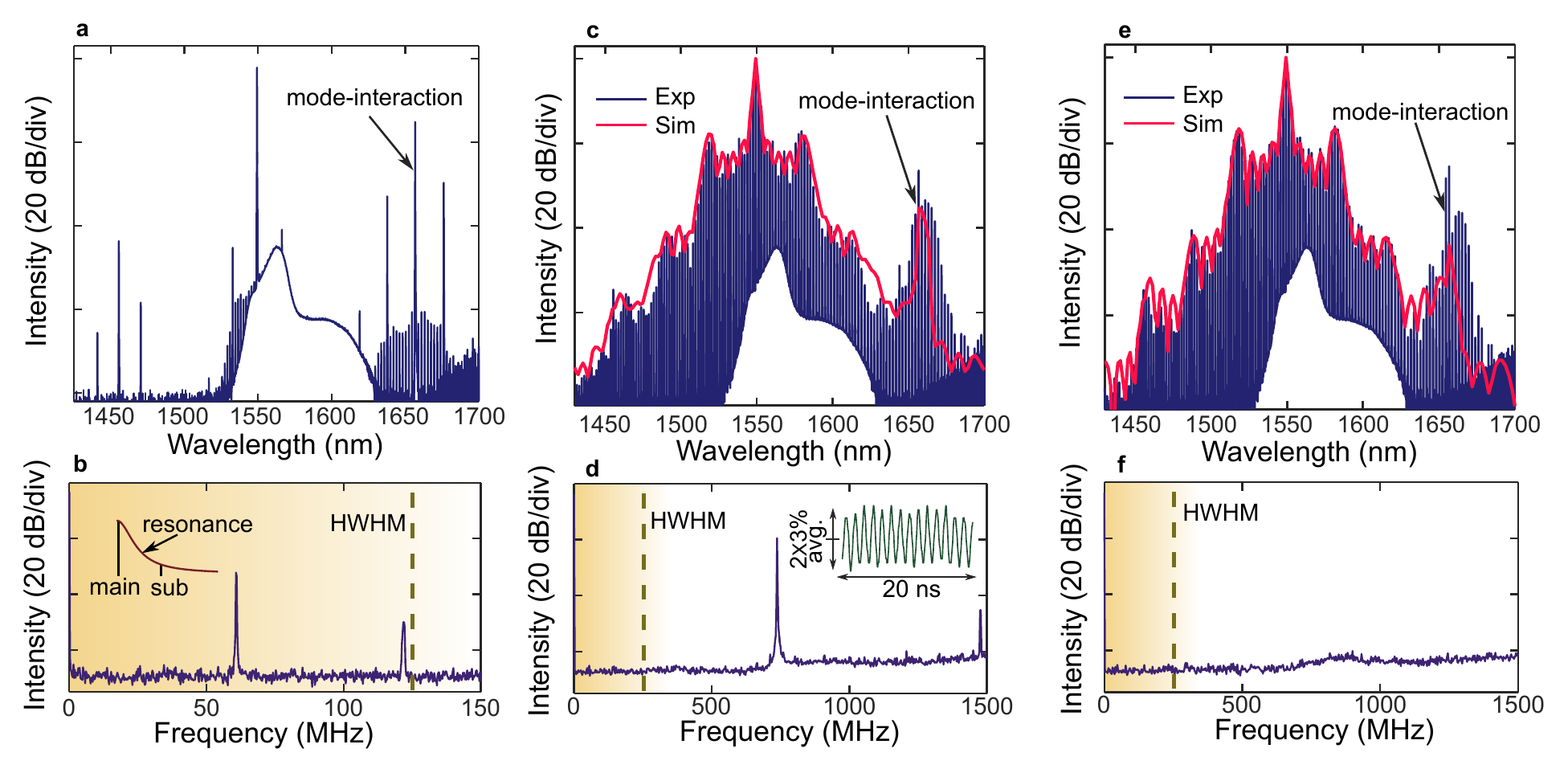}
\caption{\textbf{Generation dynamics of the dark breather.} When tuning the laser into cavity resonance, the comb is initiated by mode-interaction. \textbf{a,~b,} In the modulation instability regime, some coarsely spaced comb lines are generated and there are narrow RF peaks, whose frequencies are within the half-width-half-maximum (HWHM) of the cavity resonance. \textbf{c,~d}, With further tuning into resonance, we can get a broadband comb and a dark breather state. The simulated optical spectrum is in close agreement with the measured spectrum. The dark breather state has narrow RF peaks on the RF spectrum, whose frequencies are outside the HWHM of the resonance. The inset in \textbf{d} shows the measured comb power change of the dark breather. The dark breather only breathes weakly with a depth $\sim$3\%. \textbf{e,~f,} By lowering the pump power at the same wavelength, we can get a coherent dark pulse comb with low RF noise. The simulated optical spectrum also closely matches the measured spectrum. There is no significant difference in the optical spectra in the dark breather state and the dark pulse state.}
\label{Fig1Generation}
\end{figure*}

From a practical perspective, studying the dark breather gives insights into dark pulse Kerr comb generation in normal dispersion microresonators. The soliton Kerr combs generated in anomalous dispersion microresonators have been used in many applications, including communications \cite{Koos_Nature2017}, dual-comb spectroscopies \cite{Vahala_Science2016}, and low noise microwave generation \cite{Matsko_NC2015}. However, the conversion efficiency from the pump into the generated comb lines is usually less than a few percent for single-soliton combs \cite{Willner_OL2014,Weiner_OE2016}. In contrast, dark pulse Kerr combs can reach much higher conversion efficiency (above 30\%) \cite{Weiner_LPR2017}, providing an important tool for high power Kerr comb generation. Dark breathers represent an intrinsic source of instability for dark pulses; characterization and modeling of such breathers may  offer better opportunities for their control, and trigger the search for dark breathers in other physical systems.

\section{Results}
\subsection{Dark breather generation}

\begin{figure*}[t]
\centering
\includegraphics[width=1.85\columnwidth]{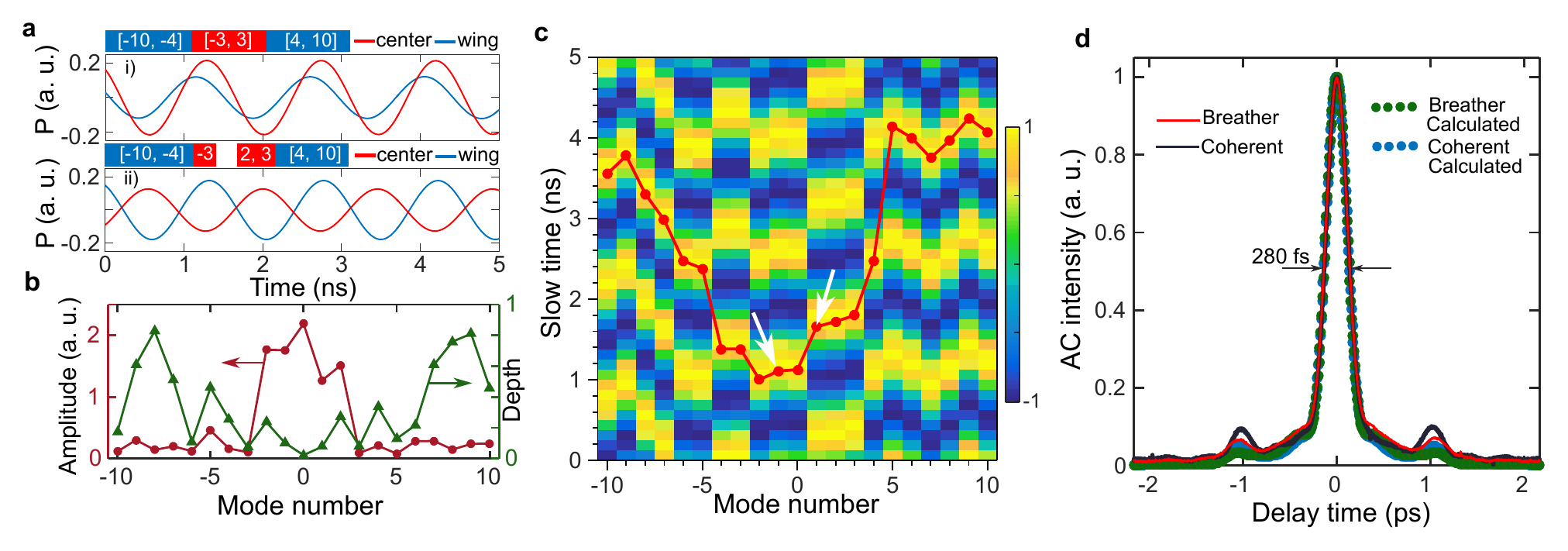}
\caption{\textbf{Characterization of the dark breather.} \textbf{a,} The dark breather exhibits energy exchange between different comb lines. When choosing the comb lines with mode numbers in $[-3, 3]$ as the center group and comb lines in $[-10, -4]\cup[4, 10]$ as the wing group, we observe a phase difference in the fast evolving spectrum, but not an anti-phase exchange in (\lowercase\expandafter{\romannumeral1}). When eliminating modes in $[-2, 1]$ for the center group (keeping mode $-3$, 2, 3 only), there is a nearly out of phase energy exchange between the wing group and the modified center group in (\lowercase\expandafter{\romannumeral2}). \textbf{b,} The absolute and fractional breathing amplitudes of individual comb lines. The absolute breathing amplitude is stronger at the center, while the breathing depth is larger at the wing. \textbf{c,} Different comb lines breathe with different phase; the change of the breathing phase varies abruptly and asymmetric with respect to the pump. The colour density shows the normalized breathing of each individual line (normalized to the breathing amplitude of the same individual lines) and the circles on line mark the peaks of the oscillation for different lines, illustrating the breathing phase. The arrows illustrate the peaks for the modes $\pm$1. \textbf{d,} The dark breather retains a dark-pulse-like waveform. By line-by-line pulse shaping, dark pulses can be shaped into transform-limited pulses. Using the same phase on the pulse-shaper, the dark breather comb can also be shaped into a transform-limited pulse.}
\label{Fig2Characterization}
\end{figure*}

The dark breather is generated in a SiN microresonator with a radius of 100 $\mu$m, a loaded $Q$-factor of 0.8$\times$10$^6$, geometry of 2000 nm$\times$600 nm, and a normal dispersion of $\beta_2$=190 ps$^2$/km. The same microresonator pumped at the same resonance was previously used to generate a mode-locked dark pulse \cite{Weiner_NP2015mode}. By tuning the cw pump laser into resonance from blue to red, we are able to initialize the comb generation via mode-interaction \cite{Weiner_NP2015mode,Weiner_Optica2014} at the pump power of 1.7 W (off-chip power, the fiber-to-chip coupling loss is $\sim$2 dB per facet). At the pump wavelength of 1549.19 nm, we get some coarsely spaced comb lines as shown in
Fig. \ref{Fig1Generation}a, which is dominated by the mode-interaction assisted modulation instability dynamics. The RF spectrum of this comb shows some narrow sidebands at low frequencies (60 MHz) (Fig. \ref{Fig1Generation}b). These RF sidebands in the modulation instability regime can be attributed to the beat note between the main comb lines and the subcomb lines (see the inset of Fig. \ref{Fig1Generation}b) as reported in ref. \cite{Kippenberg_NP2012}. Hence, the sidebands are accommodated within the halfwidth-half-maximum (HWHM) of the cavity resonance (125 MHz). Thus, the comb in Figs. \ref{Fig1Generation}a, b is not a dark breather, despite the narrow RF sidebands. When further tuning the pump into the resonance, we can get a broadband, single free-spectral-range (FSR) comb at pump wavelength of 1549.26 nm (Fig. \ref{Fig1Generation}c). The comb in this state also has narrow RF sidebands as shown in Fig. 1d. However, the breathing frequency (740 MHz, $>$12 times of the frequency in Fig. 1b) is outside the HWHM of the resonance. Thus, these sidebands are attributed to a different generation mechanism from the main-sub comb beat. Supplemented with further evidence described below, we confirm that the comb in Figs. \ref{Fig1Generation}c, d corresponds to a dark breather and attribute the sidebands to periodic breathing (modulation) of the comb lines. The averaged simulated optical spectrum (see the section on numerical simulations and Methods below) of the dark breather is in close agreement with the measured optical spectrum. An important feature of the dark breather is that it only breathes weakly. We define the breathing depth as $(P_{max}-P_{min})/(P_{max}+P_{min})$, where $P_{max(min)}$ is the maximum (minimum) power. The comb power including (excluding) the pump breathes with a depth of $\sim$3$\%$ ($\sim$5$\%$). This is significantly different from bright soliton breathers, which can breathe with a depth of $\sim$50$\%$ (excluding the pump) \cite{Weiner_PRL2016}. We can get the coherent dark pulse comb by reducing the pump power to 1.6 W at the same wavelength. Now the comb shows low-noise operation (Figs. \ref{Fig1Generation}e, f). Again, the simulated dark pulse has an optical spectrum which closely matches the measurement. There is no significant difference between the dark breather and the mode-locked dark pulse in the optical spectrum analyzer averaged spectra, as the dark breather only breathes weakly; whereas the bright soliton and bright soliton breather combs show distinct averaged spectra \cite{Weiner_PRL2016,Kippenberg_arXiv2016Breather}.

\subsection{Characterization of dark breathers}

The high frequency RF sidebands, which were also reported in \cite{Weiner_NP2015mode}, are not sufficient by themselves to confirm the observation of dark breathers. For example, a comb with a constant spectrum shape but varying power can show similar peaks. Hence, we first need to confirm there is internal energy exchange between the comb lines. To do that, we use a pulse-shaper \cite{Weiner_RSI2000} to programmably select groups of comb line and record their fast evolution \cite{Weiner_PRL2016} (see Methods and Supplementary Information Sec. 1). We first choose the comb lines whose mode number with respect to the pump are within $[-3,~3]$ (pump line mode 0 included and larger number means higher frequency) as the center group, while the comb lines within $[-10,~-4] \bigcup [4,~10]$ as the wing group. The oscillation of the recorded power change of the center group and the wing group show a phase difference of 0.21$\pi$ in Fig. \ref{Fig2Characterization}a\lowercase\expandafter{\romannumeral1}. When further suppressing the comb line within $[-2, 1]$ for the center group (i.e., only keeping lines $-$3, 2 and 3), the center group breathes out of phase (0.93$\pi$) with the wing group as shown in Fig. \ref{Fig2Characterization}a\lowercase\expandafter{\romannumeral2}. Thus, it is confirmed that there is energy exchange within the Kerr comb in Fig. \ref{Fig1Generation}c. To gain more insight into the breathing dynamics of the dark breather, the pulse shaper is programmed to select single comb lines to record their fast breathing. The absolute breathing amplitude of the central comb lines within $[-2,~2]$ is much higher compared to other comb lines. However, the average power is also highest in the central comb lines. The modulation depth of individual lines tends to be largest in the wing of the spectrum (Fig. \ref{Fig2Characterization}b). Moreover, different lines breathe with different phases as shown in Fig. \ref{Fig2Characterization}c, which differs from the breathers in the NLSE with normal dispersion \cite{Akhmediev_PRA1993} (Supplementary Section 4). Abrupt changes in the phase and amplitude of the breathing can be seen between some adjacent lines, whereas bright soliton breathers have relatively smooth changes in breathing phase \cite{Weiner_PRL2016}. We also note that the breathing phase is not symmetric with respect to the pump (e.g., see the modes $\pm$1, illustrated by the arrows). Mode-interactions may be responsible for this asymmetry, as the dominant mode-interaction only exists on one side of the pump, breaking the symmetry (see numerical simulations below).

To confirm the observation of dark breathers, we need to verify that the comb in Figs. \ref{Fig1Generation}(c, d) has a dark-pulse like waveform. We use line-by-line pulse shaping (see Methods) to probe the corresponding waveform \cite{Weiner_RSI2000,Weiner_NP2015mode,Weiner_NP2011}. We first adjust the pump power to reach the mode-locked dark pulse state and apply spectral phase shaping to obtain a transform-limited pulse at the output (Fig. \ref{Fig2Characterization}d). We then adjust the pump power to transition into the dark breather state, while keeping the same phase profile on the pulse-shaper. The output is still shaped into a transform-limited pulse. This comparison suggests that the dark breather has a spectral phase profile close to that of the mode-locked dark pulse. Since the dark breather exhibits only a gentle modulation and the power spectrum remains close to that of the dark-pulse comb, we deduce that the dark breather retains a dark-pulse-like waveform (Supplementary Section 2). These findings confirm that we are observing a dark breather in our normal dispersion microresonator.

\subsection{Transition to chaotic breathing}
\begin{figure*}[t]
\centering
\includegraphics[width=1.8\columnwidth]{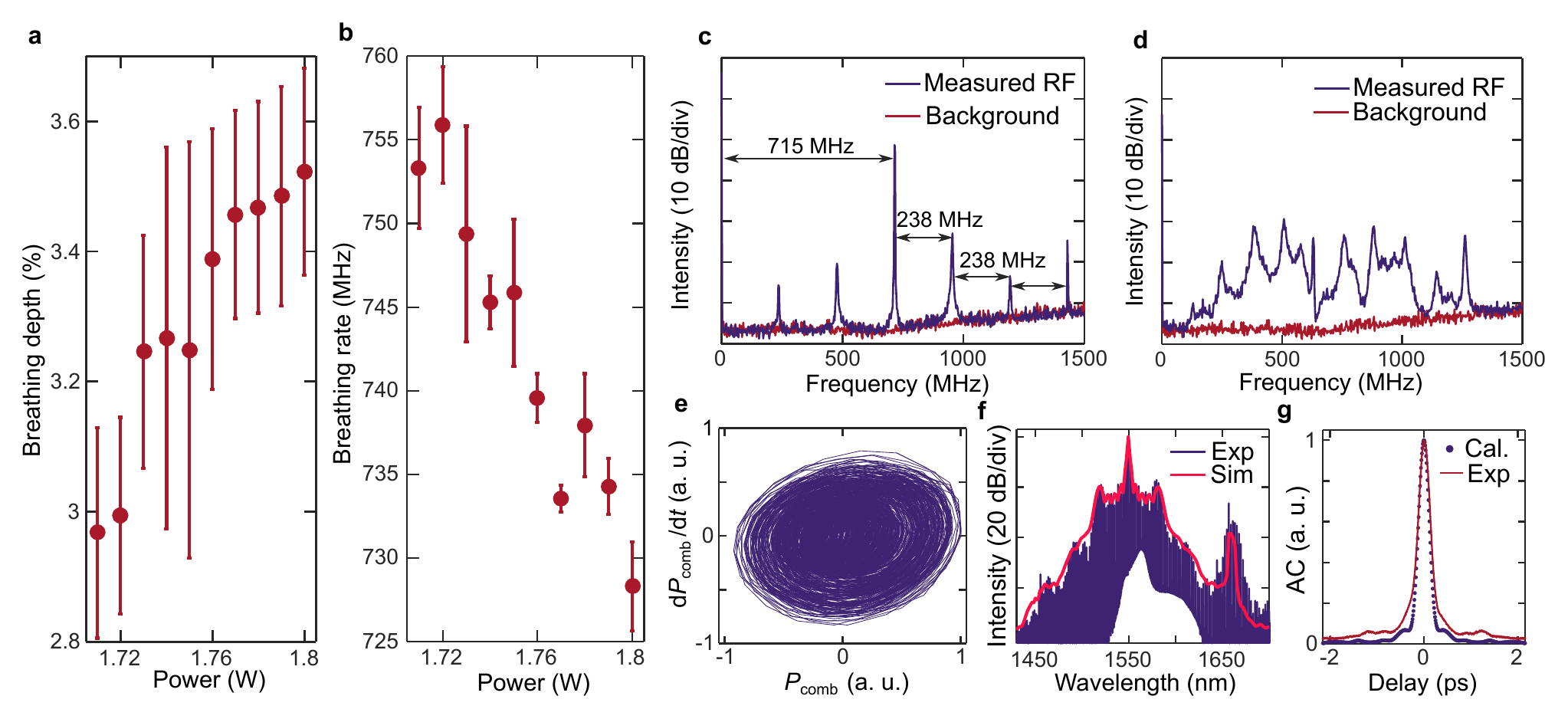}
\caption{\textbf{Power dependence of the breathing properties.} \textbf{a,} The breathing depth of the comb power increases with increasing pump. \textbf{b,} The breathing rate decreases slightly with pump power. \textbf{c,} When the pump power is increased to 1.9 W, the system transitions to a period-tripling state, as evidenced by the additional sidebands spaced by $\sim$238 MHz. \textbf{d,} Broadband RF noise is observed when the pump power is further increased to 2.1 W. \textbf{e,} The phase diagram of the output comb power (pump included), which is obtained by plotting the first-order time derivative of the comb power (d$P_{\text{comb}}$/d$t$) versus $P_{\text{comb}}$, also shows chaotic variation. \textbf{f,} The optical spectrum in this chaotic breathing state is similar to the dark pulse state; the averaged simulated spectrum in the chaotic breathing state is in close agreement with the measured spectrum. \textbf{g,} The autocorrelation of the Kerr comb in the chaotic breathing state after pulse shaping (the same amplitude and phase adjustment used in Fig. \ref{Fig2Characterization}d) is close to the calculated autocorrelation trace of the transform limited pulse. The auto-correlation of the Kerr comb in the chaotic breathing state after pulse shaping (the same amplitude and phase adjustment used in Fig. \ref{Fig2Characterization}d) is close to the calculated auto-correlation trace of the transform limited pulse.}
\label{Fig3Chaotic}
\end{figure*}

\begin{figure*}[t]
\centering
\includegraphics[width=1.9\columnwidth]{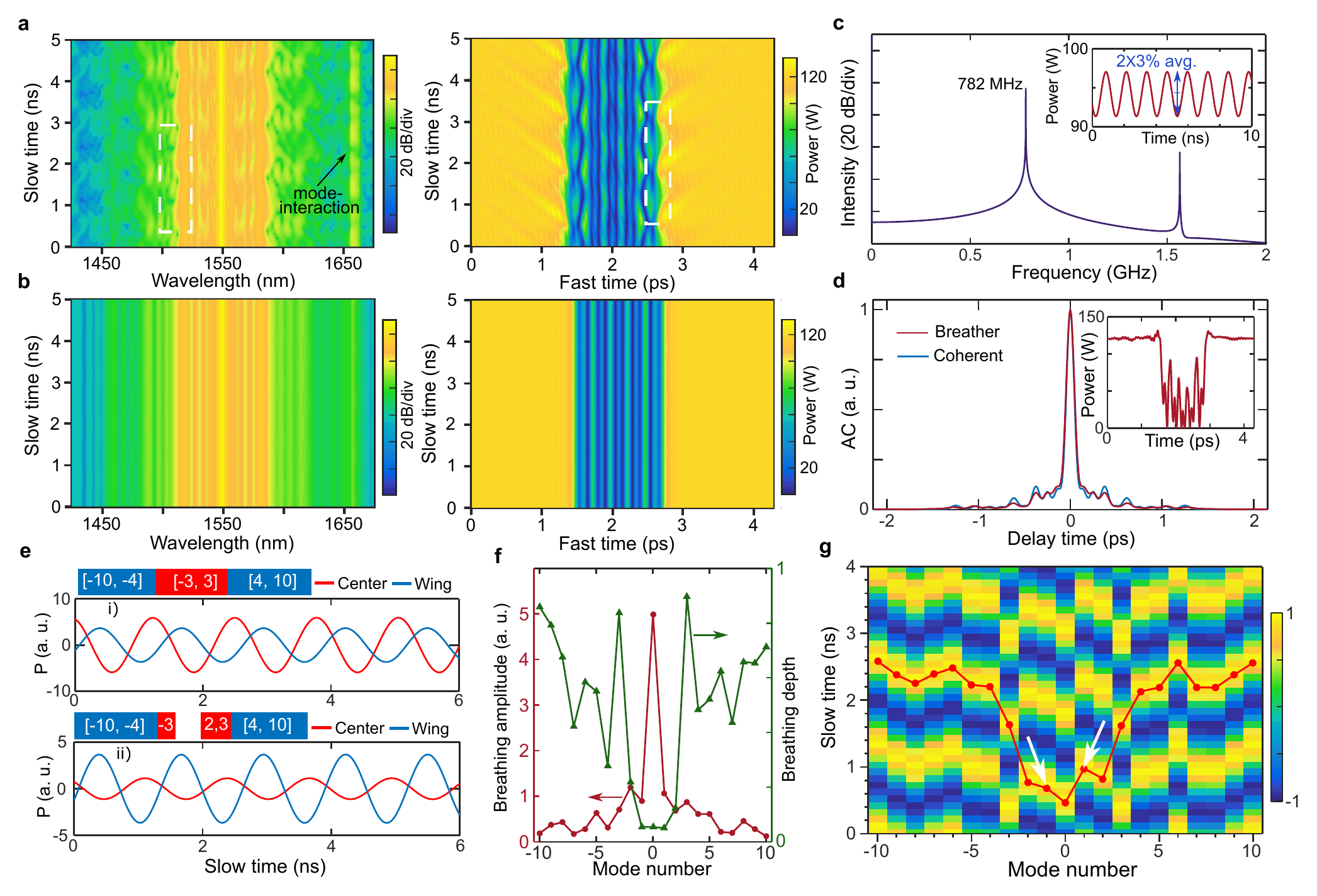}
\caption{\textbf{Numerical simulation on the dark breather dynamics.} \textbf{a,} Simulated breathing dynamics in the frequency and time domain. Mode-interaction is included in the simulation for modes in $[-57,~-54]$, leading to enhanced intensity for the perturbed modes. The dashed boxes show the breathing of the spectrum and the waveform. \textbf{b,}  Simulated spectral and temporal dynamics of the coherent dark pulse comb when decreasing the pump power to 628 mW. \textbf{c,} Simulated RF spectrum of the dark breather state, showing a breathing rate of 782 MHz, close to the experiments. The inset shows the change of the total comb power including pump; the breathing depth is about 3$\%$. \textbf{d,} The averaged simulated autocorrelation trace of the pulse-shaped dark breather comb (see Methods for numerical shaping method) is close to the autocorrelation of the pulse-shaped dark pulse comb. The inset is an example of the breathing dark-pulse-like waveform. \textbf{e,} The simulated energy exchange between comb lines at the center and the wing when selecting the comb line groups in the same manner with experiments in Fig. 2a. \textbf{f,} The simulated absolute breathing amplitude and fractional breathing of individual lines. \textbf{g,} The breathing phase of different comb lines. The colour density shows the normalized breathing of individual lines and the circles on the line mark the peaks of the breathing for different comb lines. The arrows illustrate that modes $\pm$1 breathe at different phases.
}
\label{Fig4simulations}
\end{figure*}

\begin{figure*}[t]
\centering
\includegraphics[width=1.9\columnwidth]{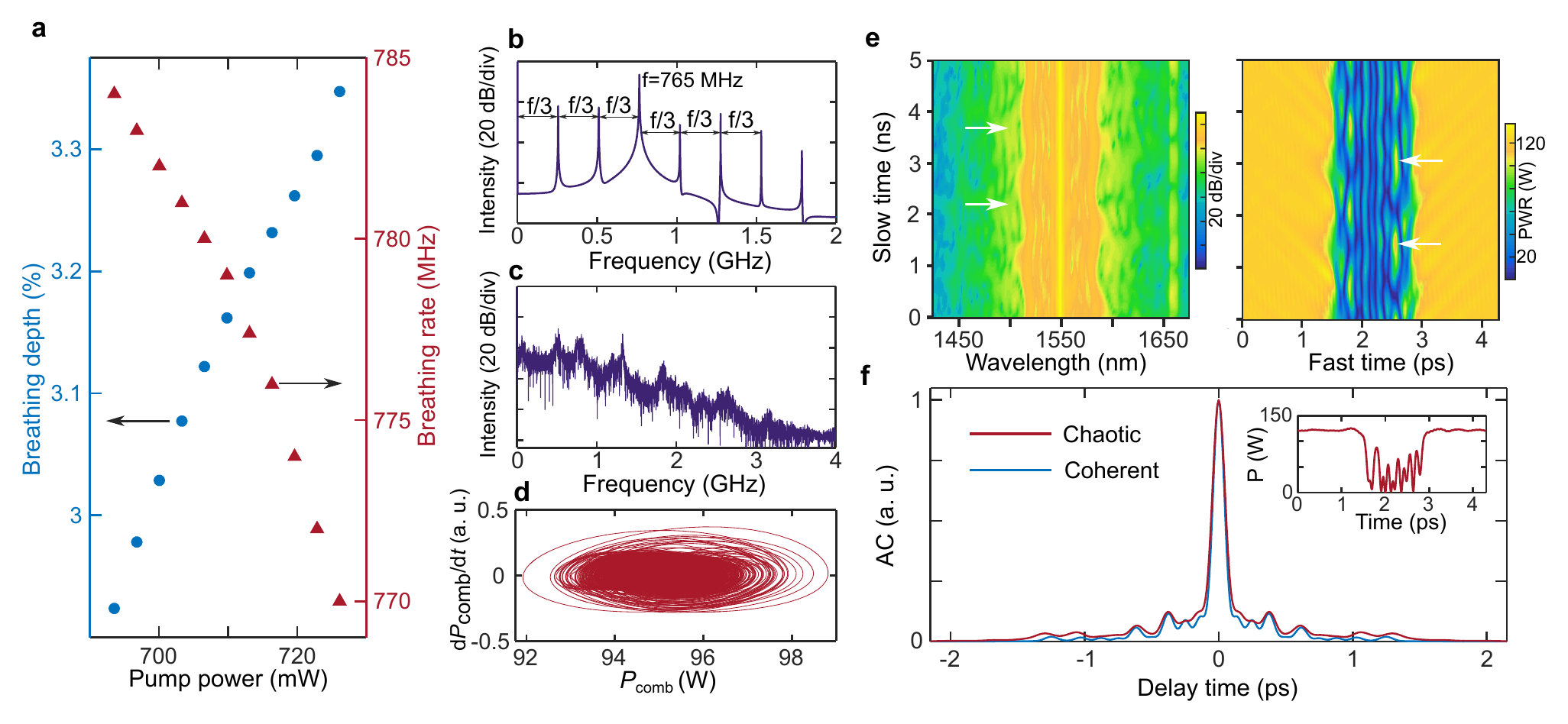}
\caption{\textbf{Numerical simulations on chaotic breathing.} \textbf{a,} When the pump power increases, the breathing depth of the total comb power increases slightly and the breathing rate decreases slightly. \textbf{b,} When the pump power is increased to 752 mW, the dark breather goes through a tripling of the breathing period and the RF spectrum shows two equidistantly spaced weak tones between 0 and the strongest tone at 765 MHz. \textbf{c,} When further increasing the pump power to 785 mW, the breathing becomes chaotic and the RF spectrum becomes broadband, losing the original narrow sidebands. \textbf{d,} An example of the total comb power (including the pump line) plotted against its time derivative in the chaotic breathing state, showing stochastic variation. \textbf{e,} The spectral and temporal breathing dynamics. The arrows highlight the nonperiodic spectrum and the waveform evolution. \textbf{f,} When numerically shaping the chaotic breathing comb into a pulse (see Methods), the averaged auto-correlation trace is close to the auto-correlation of the shaped coherent comb. The inset depicts an example of the waveform of the chaotic breathing state, showing the dark-pulse-like waveform is retained.}
\label{Fig5simulations}
\end{figure*}

Fig. \ref{Fig1Generation} shows the breathing instability can be eliminated by decreasing the pump power. On the other hand, the absolute breathing depth increases slightly when increasing the pump power (Fig. \ref{Fig3Chaotic}a). At the same time, the breathing rate decreases slightly (Fig. \ref{Fig3Chaotic}b). The breathing rate of bright soliton breathers in the anomalous dispersion regime also decreases with increasing pump power \cite{Kippenberg_arXiv2016Breather}. The decrease in the breathing rate is an indication of approaching critical transitions for dynamical systems \cite{Kippenberg_arXiv2016Breather,Sugihara_Nature2009}. When increasing the pump power further to 1.9 W, the breathing of the dark breather goes through a period-tripling bifurcation (Fig. \ref{Fig3Chaotic}c). The period-doubling bifurcation occurs more commonly in dynamical systems \cite{Ott_2002} including bright soliton breathers in microresonators \cite{Gaeta_NC2017,Kippenberg_arXiv2016Breather}; however, it is not observed for the studied dark breather. The perturbation from mode-interaction is found to be important to this period-tripling, without which it does not occur (see Supplementary Section 5). Such a direct period-tripling transition was also demonstrated in passively mode-locked fiber lasers \cite{Akhmediev_PRE2004}.

By increasing the pump power further to 2.1 W, the breathing becomes chaotic. The measured RF spectrum shows broadband noise (Fig. \ref{Fig3Chaotic}d), and the phase diagram (plotting first-order time derivative of comb power d$P_{comb}$/d$t$ versus $P_{comb}$) shows the comb power changes chaotically (Fig. \ref{Fig3Chaotic}e). Chaotic breathing was predicted and observed (in fibre cavities) to exist in the LLE with anomalous dispersion \cite{Bekki1985,Coen_OE2013,Coen_Optica2016}; it has also been predicted to exist for dark breathers but not demonstrated yet \cite{Gelens_PRA2016}. Our results constitute the first observation of such a chaotic transition for dark breathers in all physical systems, to our knowledge. Note this chaotic breathing state is distinct from the chaotic state prior to the mode-locking transition \cite{Weiner_NP2015mode,Kippenberg_NP2014temporal} (see Supplementary Sections 1, 3). The comb in the chaotic breathing state retains the optical spectral features of the dark pulse (Fig. \ref{Fig3Chaotic}f), and the averaged spectrum in simulations is also in good agreement with the measurement. When we apply the same phase profile used in Fig. \ref{Fig2Characterization}d on the pulse shaper, the auto-correlation (AC) trace obtained under the chaotic breathing state still shows a clean pulse which is close to the transform-limit (as calculated from the spectrometer averaged spectrum). The visibility of the AC trace exceeds 97\% of the expected value; this differs dramatically from what is observed with the chaotic comb in the absence of a mode-locking transition \cite{Weiner_NP2011}. The AC results suggest the chaotic breathing state has a dark-pulse-like waveform (see also the simulation results in Fig. \ref{Fig5simulations}).  This is consistent with the fact that excitation of the chaotic breathing is reversible, i.e., the dark breather and the coherent comb can be obtained by simply decreasing the pump power from this chaotic state.

\subsection{Numerical simulations}

Kerr comb generation dynamics can be modeled by the LLE, i.e., the driven-damped NLSE \cite{Lugiato_PRL1987spatial,Coen_OE2013,Weiner_PRL2016}. For the NLSE, breather-like solutions also exist with normal dispersion \cite{Akhmediev_PRA1993} (Supplementary Section 4). However, dissipative effects in externally pumped microresonators change the dark breather dynamics dramatically. Here, we use the LLE to model the dark breather dynamics. When choosing the pump power as 700 mW and detuning as 0.0629, a dark breather can be generated (see Methods and Supplementary Section 3). As shown above, the simulated spectrum (averaged over slow time) is in close agreement of the measured spectrum (Fig. \ref{Fig1Generation}c). The inclusion of the mode-interaction gives the long wavelength peak around 1657 nm. From Fig. \ref{Fig4simulations}a, we can see the periodic variation of the spectral and temporal shape of the dark breather (see the dashed boxes). From the temporal breathing, we can see the waveform-background (high power part of the dark-pulse-like waveform) stays nearly unchanged in the breathing and the waveform-hole (low power part of the dark-pulse-like waveform) breathes more strongly (Supplementary Animation 1). This helps to elucidate the low breathing depth of the comb power, as the waveform-top carries most of the power. With the same detuning, a coherent dark pulse comb can be obtained by decreasing the pump power to 628 mW (Fig. \ref{Fig4simulations}b), consistent with experiments. As shown in Fig. \ref{Fig4simulations}c, the simulated dark breather has a breathing rate of 782 MHz, and the breathing depth of the comb power including (excluding) the pump is 3\% (6\%), both in agreement with the experiments. The simulation retains the dark-pulse-like waveform in the breathing state; the simulated AC (averaged over slow time, see Methods) remains close to the AC of the coherent comb after numerical pulse shaping (Fig. \ref{Fig4simulations}d).

The simulated dark breather also exhibits energy exchange between comb lines. When choosing the comb lines in the same manner as used in Fig. \ref{Fig2Characterization}a, the simulated power change is shown in Fig. \ref{Fig4simulations}e. The center group in $[-3, 3]$ has a different breathing phase with the wing group in $[-10, -4]\cup[4, 10]$, but is not completely out of phase (0.70$\pi$). Similar to experiments, when eliminating modes in $[-2, 1]$ from the center group, the center group and the wing group exhibit nearly out of phase energy exchange (1.11$\pi$). The breathing behaviour also differs greatly from line to line. The absolute breathing amplitude is stronger near the spectrum center, while the breathing depth tends to be stronger at the wing (Fig. \ref{Fig4simulations}f). The discrepancy of the absolute breathing amplitude from experiments (Fig. \ref{Fig2Characterization}b) may be attributed to the difficulty in modeling the mode-interaction accurately. The simulated breathing amplitude is also not symmetric with respect to the pump. The breathing phase of different lines changes asymmetrically with respect to the pump (see especially modes $\pm1$ in Fig. \ref{Fig4simulations}(g)). Note that the breathing becomes symmetric with respect to the pump when mode-interaction is excluded (Supplementary Section 5). This simulation together with the experimental result in Fig. \ref{Fig2Characterization}c suggests that the comb lines in the dark breathers are highly coupled and the dark pulse instability can be affected by a perturbation over 12 THz away. Furthermore, the breathing phase can change abruptly between adjacent lines, in agreement with experiments.

The simulated dark dynamics also depend strongly on the pump power. In agreement with experiments, the simulated breathing depth increases slightly while the breathing rate decreases slightly with increasing pump power, as shown in Fig. \ref{Fig5simulations}a. Moreover, when the pump power is further increased, the simulated dark breather undergoes a transition into chaotic breathing via period-tripling. When increasing the pump power to 752 mW, the simulated RF spectrum shows a period-tripling feature (Fig. \ref{Fig5simulations}b). Note that this period-tripling does not occur in the absence of mode-interaction (see Supplementary Information Sec. 5). When further increasing the pump power to 785 mW, the RF spectrum becomes broadband (Fig. \ref{Fig5simulations}c). In addition, the phase diagram of the total comb power shows chaotic features (Fig. \ref{Fig5simulations}d). The simulated spectral and temporal evolution are shown in Fig. \ref{Fig5simulations}e. From the arrows therein, we can see the breathing periodicity is lost (Supplementary Animation 2); however, the dark-pulse-like waveform and the dark-pulse-featured spectrum are retained. Furthermore, the averaged AC trace (after phase compensation) of the chaotic breathing state remains close to the AC of the coherent state. Breathers in the frame of NLSE cannot undergo period-bifurcation or chaotic transition, as the NLSE is fully integrable \cite{Manakov1974}. Here, the dissipative effects of the microresonators enable the observation of this chaotic transition. These transitions should also be possible for dark breathers in other systems, e.g., spatial solitons \cite{Lugiato_PRL1987spatial} and plasmas \cite{Bekki1985}.

\section{Discussion and conclusion}
Optical dark breathers in normal dispersion microresonators are clearly observed and comprehensively modeled for the first time. The dark breather features a high frequency modulation, a weak breathing depth, and an energy exchange between different lines. The breathing dynamics of comb lines around the pump are shown to be affected a mode-interaction over 12 THz away, which reveals the influence of mode-interaction on the instability of mode-locked dark pulses. A transition from periodic breathing into a chaotic breathing state where the comb retains a dark-pulse-like waveform is also observed, which highlights the dissipative features of the breathers in cavities. Since dissipation is unavoidable in the real world, we anticipate more features imprinted on breathers by dissipation can be revealed in optics and other physical systems. Our work constitutes an important addition to the study of breathers and may have implications to a variety of physical systems, including hydrodynamics, plasmas, BEC and granular lattices.

\section{Method}
\small\textbf{Fast recording of the spectral evolution.} The method to record the fast evolution of the spectral breathing is the same as that used in ref. \cite{Weiner_PRL2016} (see the Supplementary Information Sec. 1 for the experimental setup). After a pulse-shaper is used to select specific comb lines, the output is recorded by a fast photodiode. A portion of the output is used to trigger the oscilloscope in order to provide a timing reference for measurements performed on different comb lines. The recorded traces are Fourier transformed and numerically filtered by a 20 MHz bandpass filter to isolate the strongest breathing tone; the filtered spectra are inverse Fourier transformed to yield the spectral breathing traces presented in Figs. \ref{Fig2Characterization}a, c.

\textbf{Line-by-line pulse shaping.} The method for shaping the output comb into a transform-limited pulse is described in ref. \cite{Weiner_NP2011}. Comb lines with mode number in $[-7,~11]$ are used for pulse-shaping. The intensity of the comb lines is adjusted to prevent the strongest comb lines from dominating the pulse shaping. In this intensity adjustment, the pump line (mode 0) is attenuated by 20 dB; modes $\pm$1 are attenuated by 10 dB; modes $\pm$2 are attenuated by 5 dB, and the other modes are not attenuated.

\textbf{Numerical simulations.} The simulations are based on the Lugiato-Lefever equation \cite{Lugiato_PRL1987spatial,Coen_OL2013modeling}, which can be written as
\begin{equation}
\begin{aligned}
\left( {{t}_{R}}\frac{\partial }{\partial t}+\frac{\alpha +\kappa }{2}+i\frac{{{\beta }_{2}}L}{2}\frac{{{\partial }^{2}}}{\partial {{\tau }^{2}}}+i{{\delta }_{0}}-i\gamma L|E|^2 \right)E=\sqrt{\kappa P_{in}}
\end{aligned}
\label{eq:LLE}
\end{equation}
where $E$, $t_R$, $L$, $\beta_2$, $\gamma$, $\alpha$, $\kappa$, $\delta_0$, and $P_{in}$ are the envelope of the intracavity field, round-trip time, cavity length, group-velocity dispersion, nonlinear coefficient, intrinsic loss, coupling loss, pump phase detuning and pump power, respectively. In the simulations, the parameters used are $t_R$=4.3 ps, $\beta_2$=187 ps$^2$/km, $L$=628 $\mu$m, $\gamma$=0.9 (Wm)$^{-1}$, $\alpha$=0.0014, $\kappa$=0.0054, $P_{in}$=700 mW, which are representative of the studied device. Mode-interaction is included in the simulation by adding a phase shift $\Delta \phi$ per round-trip to the perturbed modes. In simulations, 4 modes within $[-57,~-54]$ are perturbed with an identical $\Delta \phi$. In the first step (0-100 ns), for which $\delta_0$=0.0102 and $\Delta \phi=-$0.815, the comb generation is initiated by the mode-interaction. In the second step (100-200 ns), for which $\delta_0$=0.0272, $\Delta \phi=-$0.815, the comb grows and becomes chaotic. In the third step (200-300 ns), for which $\delta_0$=0.0629, $\Delta \phi=-$0.326, a dark breather can be obtained (see Supplementary Information Sec. 3 for the comb generation dynamics).

The spectral breathing is numerically filtered in the same way as experiments to obtain the traces in Figs. \ref{Fig4simulations}e, g. For the traces presented in Fig. \ref{Fig4simulations}d and Fig. \ref{Fig5simulations}f, we extract the simulated phase of the coherent dark pulse, which we denote $\phi_{\text{dp}}(\omega)$, and then add $-\phi_{\text{dp}}(\omega)$ to compensate the phase and obtain a transform-limited pulse.  The AC is then calculated.  For the dark breather and chaotic breathing comb, the same $-\phi_{\text{dp}}(\omega)$is added to the complex spectral amplitude. Then a series of AC traces are computed based on snapshots of the intensity profile. 10$^4$ individual AC traces are calculated in a 1 $\mu$s time window (equivalent to $\sim$782 periods for the dark breather) and averaged to yield the traces in Fig. \ref{Fig4simulations}d and Fig. \ref{Fig5simulations}f. All the simulated comb modes are used in the simulated pulse shaping; the intensity of the comb lines within $[-2, ~2]$ are adjusted in the same way used in the line-by-line pulse shaping in experiments.
~~~

\textbf{Acknowledgements.} This work was supported in part by the Air Force Office of Scientific Research (AFOSR) (Grant No. FA9550-15-1-0211), by the DARPA PULSE program (Grant No. W31P40-13-1-0018) from AMRDEC, and by the National Science Foundation (NSF) (Grant No. ECCS-1509578). We gratefully acknowledges fruitful discussions with Xiaoxiao Xue and Nail Akhmediev.




\bibliography{reflist}
\end{document}



\title{Supplementary Information for ``Dark breathers in a normal dispersion microresonator"}

\author{Chengying Bao$^1$}
\email{bao33@purdue.edu}


\author{Yi Xuan$^{1,2}$}

\author{Daniel E. Leaird$^{1}$}

\author{Minghao Qi$^{1,2}$}

\author{Andrew M. Weiner$^{1,2}$}
\email{amw@purdue.edu}
\affiliation{$^{1}$School of Electrical and Computer Engineering, Purdue University, 465 Northwestern Avenue, West Lafayette, Indiana 47907-2035, USA}
\affiliation{$^{2}$Birck Nanotechnology Center, Purdue University, 1205 West State Street, West Lafayette, Indiana 47907, USA}

\maketitle

\renewcommand{\thefigure}{S\arabic{figure}}
\renewcommand{\theequation}{S\arabic{equation}}

\renewcommand*{\citenumfont}[1]{S#1}
\renewcommand*{\bibnumfmt}[1]{[S#1]}
\renewcommand{\thesection}{\arabic{section}}

\section*{1. Experimental setup and comb states}

\begin{figure}[h]
\includegraphics[width=0.75\columnwidth]{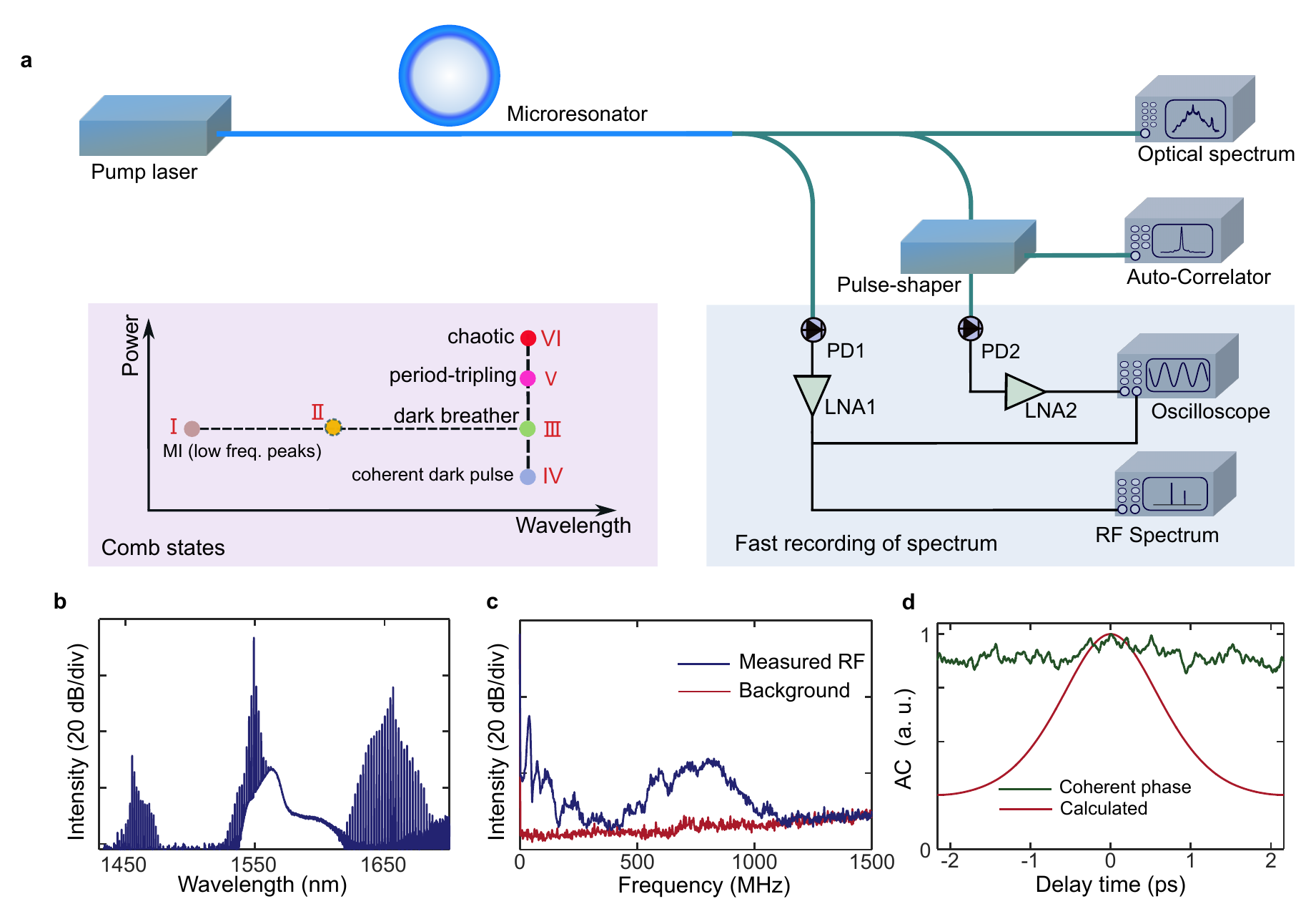}
\caption{\textbf{Experimental setup and comb states.} \textbf{a,} The Kerr comb used for dark breather study is generated from a SiN microresonator, pumped by a cw laser. To record the fast evolution associated with the comb breathing, either individual comb lines or groups of lines are selected by a pulse-shaper; their power is recorded as a function of time by PD2. A fraction of the full comb is sent to a second photodiode, PD1, which is used for synchronization between different traces. An autocorrelator and the pulse shaper are also used for the time domain characterization. The comb states presented in the main text and their corresponding pump power and wavelength are also summarized. PD: photodiode, MI: modulation instability. LNA: low noise amplifier. There is a chaotic state prior to dark breather formation, i.e., state \uppercase\expandafter{\romannumeral2}, which is distinct from the chaotic breathing state \uppercase\expandafter{\romannumeral6}. The optical spectrum and RF spectrum of this state are shown in \textbf{b} and \textbf{c} respectively. \textbf{d,}  When using the phase compensation that is used to shape the coherent comb into transform limited pulses, the comb in chaotic state II cannot be shaped into a pulse. The autocorrelation (AC) trace (green line) differs markedly from the AC of the calculated transform limited pulse.}
\label{Fig1Setup}
\end{figure}

The experimental setup and the comb states described in the main text and the Methods section are illustrated in Fig. \ref{Fig1Setup}a. The pulse shaper is used for recording of the fast spectral breathing as well as for time domain characterization. The five comb states described in the main text are also summarized: \uppercase\expandafter{\romannumeral1}) a modulation instability dynamics dominated comb with low frequency RF sidebands; \uppercase\expandafter{\romannumeral3}) tuning the pump into the resonance further generates the dark breather; \uppercase\expandafter{\romannumeral4}) by decreasing the pump power from the dark breather state, a coherent dark pulse state can be obtained; \uppercase\expandafter{\romannumeral5}) by increasing the pump power from the dark breather state, a period-tripling breathing state is obtained; \uppercase\expandafter{\romannumeral6}) by increasing the pump power further, the breathing becomes chaotic. In addition to the comb states shown in the main text, there is another chaotic state (i.e., state \uppercase\expandafter{\romannumeral2}) that is observed  prior to dark breather formation (see Fig. \ref{Fig1Setup}c for its RF spectrum), which is distinct from the chaotic state (\uppercase\expandafter{\romannumeral6}) which is reached from the dark breather state. The optical spectrum of state \uppercase\expandafter{\romannumeral2} only shows some clustered comb lines (Fig. S1b). Furthermore, this chaotic comb state cannot be shaped into transform-limited pulses \cite{Weiner_NP2011}. Here, we select 11 lines (including the pump line whose intensity is attenuated by 10 dB) around the pump for pulse shaping and apply the phase profile that is used to shape the coherent dark pulse comb into transform-limited pulse, the AC trace deviates from the calculated AC of the transform-limited pulse greatly (Fig. \ref{Fig1Setup}d).

\section*{2. Intracavity waveform retrieval}

\begin{figure}[h]
\includegraphics[width=0.75\columnwidth]{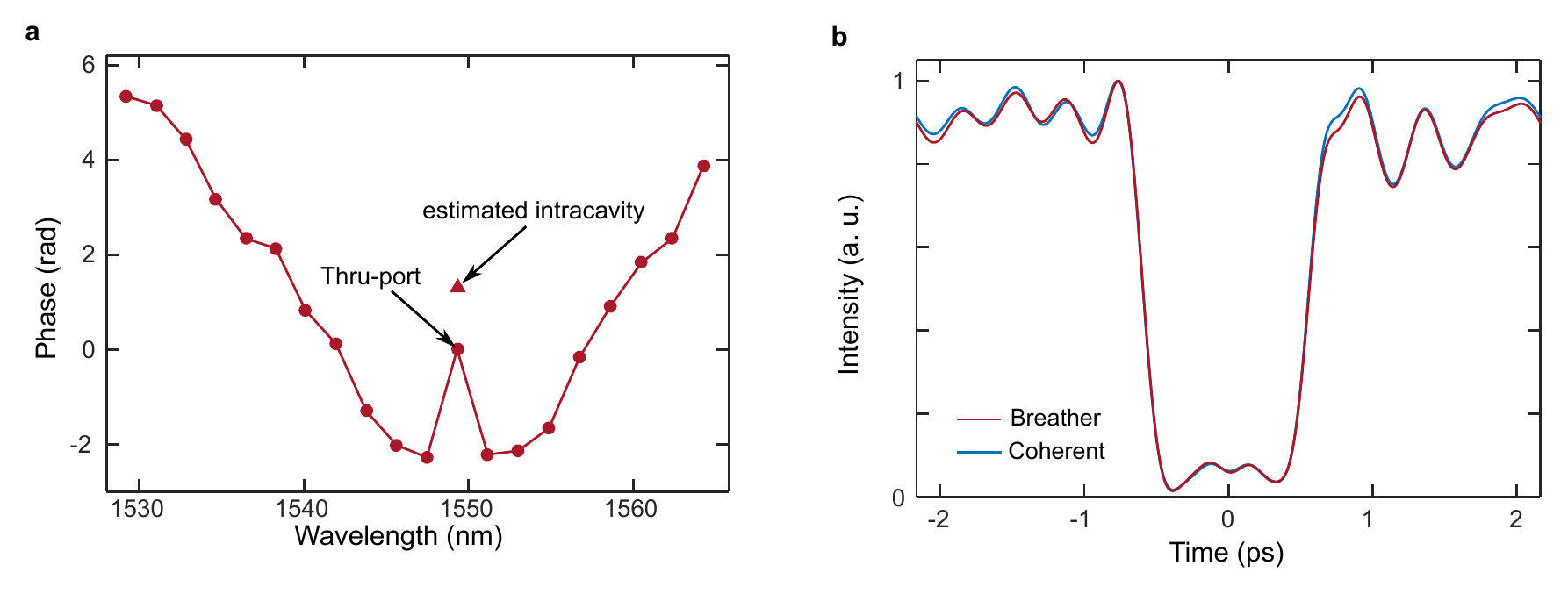}
\caption{\textbf{Intracavity waveform retrieval.} \textbf{a,} The retrieved comb phase via line-by-line pulse shaping. The intracavity pump phase is estimated based on the through-port data (see ref. \cite{Weiner_NP2015mode} and the Supplementary Information therein). \textbf{b,}  Using the averaged power spectra of the coherent dark pulse and the breather together with the retrieved phase, the intracavity waveform can be reconstructed, showing dark-pulse-like waveforms. Due to the difference in the averaged power spectra between the breather and the coherent state, the corresponding waveforms slightly differ.}
\label{Fig2Waveform}
\end{figure}

Using the comb phase retrieved in line-by-line pulse shaping, the intracavity waveform can be reconstructed \cite{Weiner_NP2015mode}. Note that the pump line in the through-port contains both the transmitted uncoupled pump and the coupled-out component from the cavity. Hence, the intensity and phase in the through-port are corrected to yield the pump line in the cavity, based on the method described in ref. \cite{Weiner_NP2015mode} and the Supplementary Information therein. By applying the retrieved phase in Fig. \ref{Fig2Waveform}(a) to the spectra of the dark breather and the coherent dark pulse, we can calculate dark-pulse-like waveforms. Since the dark breather breathes weakly, this further confirms that the dark breather retains the dark-pulse-like waveform.

\section*{3. Generation dynamics of the dark breather in simulations}

\begin{figure}[h]
\includegraphics[width=0.75\columnwidth]{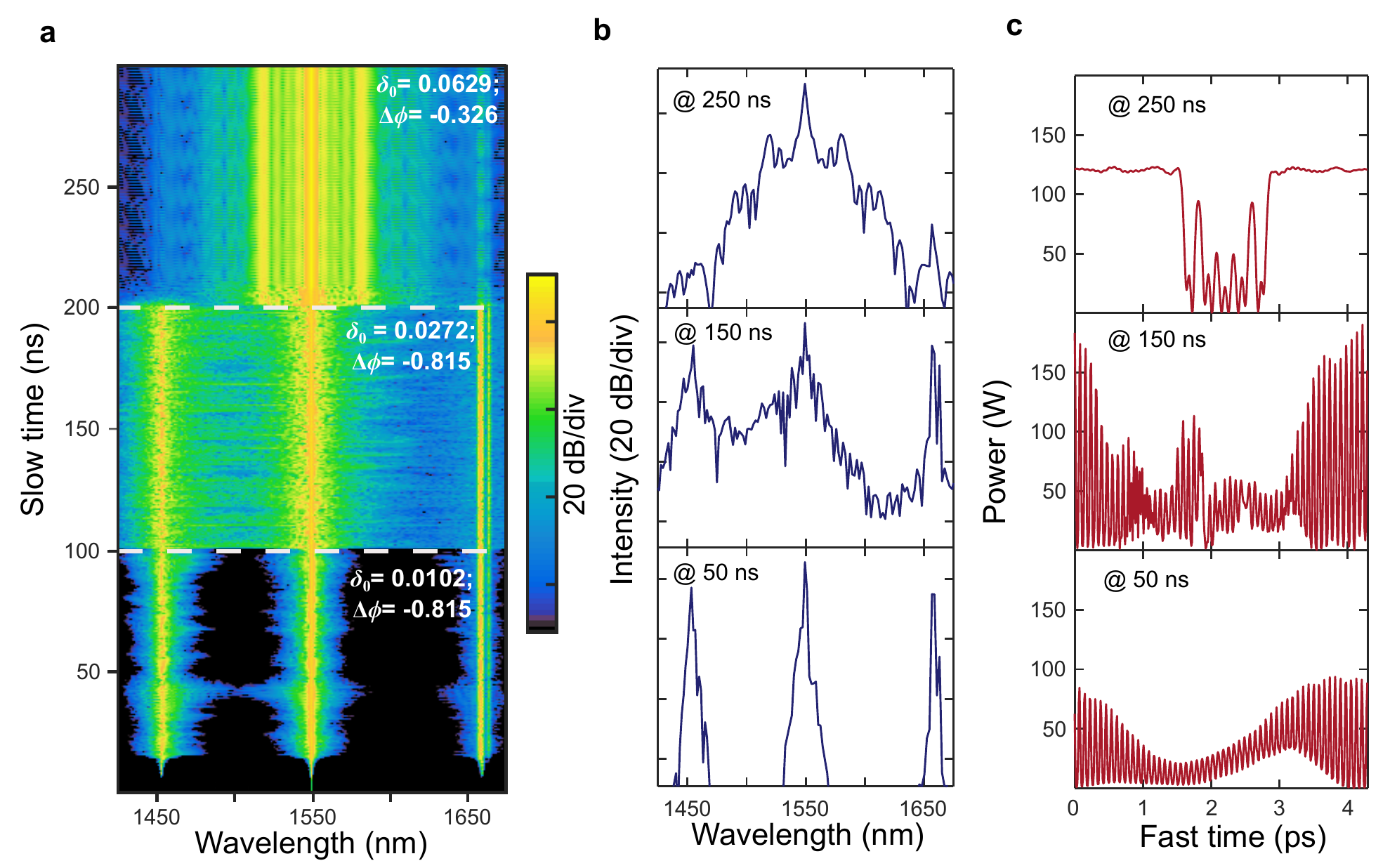}
\caption{\textbf{Tuning the detuning to generate the dark breather.} \textbf{a,}  The spectral dynamics while detuning and mode interaction parameter $\Delta \phi$ are varied. \textbf{b,}  Typical spectra at 50 ns, 150 ns and 250 ns; at 250 ns, a dark-pulse-comb featured spectrum is obtained when the dark breather is generated. \textbf{c,}  Typical waveforms at 50 ns, 150 ns and 250 ns; this further confirms that the dark breather retains the dark-pulse-like waveform.}
\label{Fig3Generation}
\end{figure}

As described in the Methods, the dark breather is generated by increasing the detuning in a step-like way in simulations. In Fig. \ref{Fig3Generation}a, we show the generation dynamics in the 3 steps. In 0$-$100 ns, $\delta_0$=0.0102, $\Delta \phi=-$0.815, and the comb is initiated by the mode-interaction and coarsely spaced. In 100$-$200 ns, $\delta_0$=0.0272, $\Delta \phi=-$0.815, and the comb grows and becomes chaotic. In 200$-$300 ns, $\delta_0$=0.0629, $\Delta \phi=-$0.326, and a dark breather can be obtained. Since the full evolution over 200$-$300 ns are plotted in Fig. \ref{Fig3Generation}a, it is hard to resolve the breathing dynamics. For visualization of the breathing on a smaller time scale, see Fig. 4 in the main text. In Figs. \ref{Fig3Generation}b, c, we show the typical spectra and waveforms (at 50 ns, 150 ns, 250 ns) in each step. Note that the chaotic comb in 100$-$200 ns (prior to dark breather formation) is different from the chaotic breathing in Fig. 3 of the main text, and the time-domain waveform and spectrum do not resemble that which occurs in the coherent dark pulse state. This state is similar to the state \uppercase\expandafter{\romannumeral2} in Fig. \ref{Fig1Setup}.

\section*{4. Breathing solution in the nonlinear Schr\"{o}dinger equation}

\begin{figure}[h]
\includegraphics[width=0.65\columnwidth]{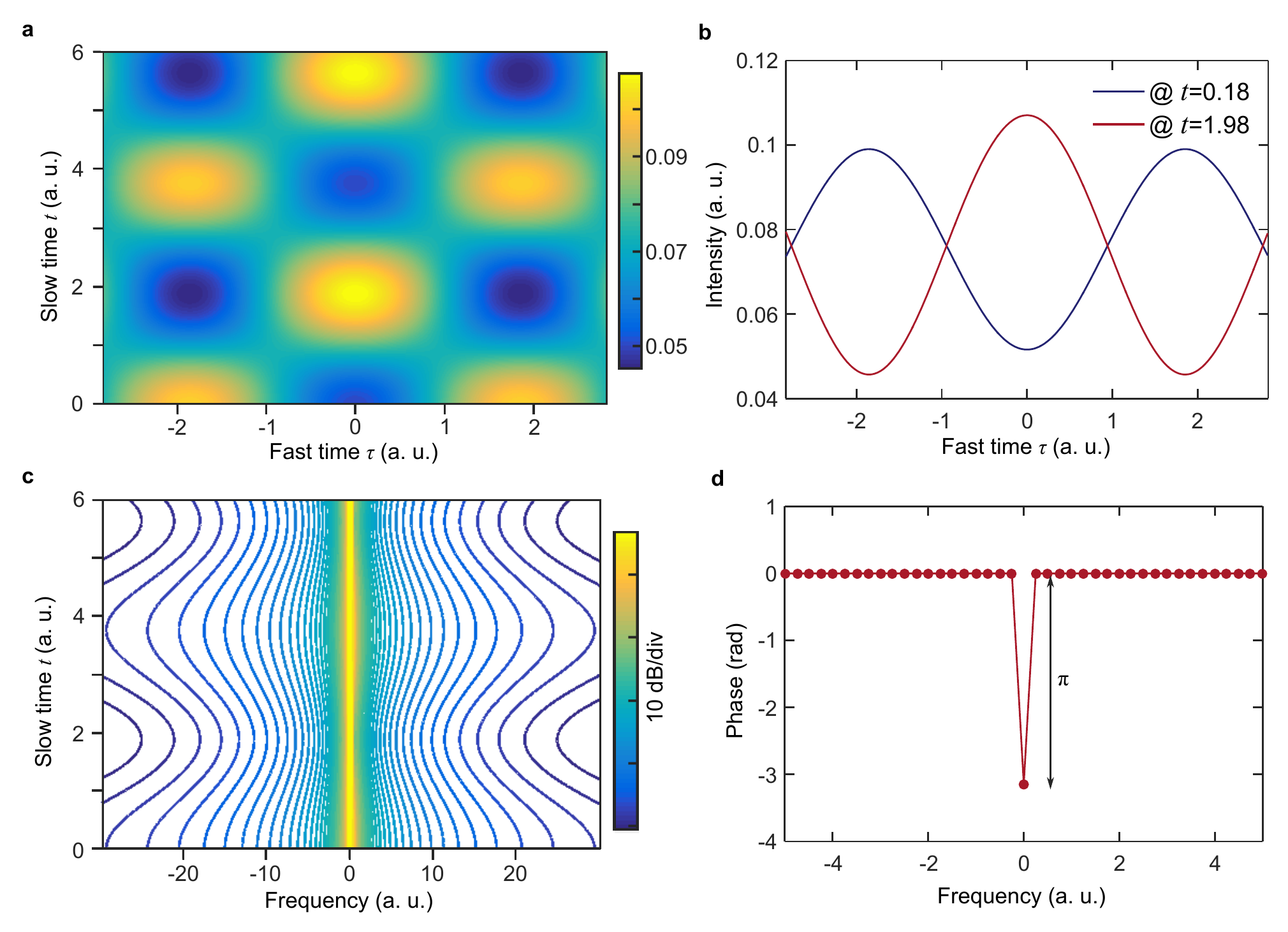}
\caption{\textbf{Breathing dynamics of the double period solution at $\epsilon$=0.5.} \textbf{a,} Temporal dynamics of the solution, \textbf{b,} two typical examples of the waveform at $t$=0.18 and $t$=1.98. \textbf{c,} Spectral breathing of the solution, \textbf{d,} all the frequency components (aside from the 0-frequency) breathe at the same phase, which is out phase of the breathing phase of the 0-frequency component.}
\label{Fig5DSP}
\end{figure}

An analytical breather solution to the nonlinear Schr\"{o}dinger equation (NLSE) also exists in the normal dispersion regime \cite{Akhmediev_PRA1993}. The breathing solution is also referred as the ``double period" solution, which can be written as \cite{Akhmediev_PRA1993}
\begin{equation}
\begin{aligned}
\Psi (\tau ,t)=\epsilon \frac{cn(t,\epsilon )-i\sqrt{1+\epsilon }sn(t,\epsilon )dn\left[ \sqrt{1+\epsilon }\tau ,\sqrt{{2\epsilon }/{(1+\epsilon )}\;} \right]}{\sqrt{1+\epsilon }dn\left[ \sqrt{1+\epsilon }\tau ,\sqrt{{2\epsilon }/{(1+\epsilon )}\;} \right]+dn(t,\epsilon )}\exp (it),
\end{aligned}
\label{eq:DPS}
\end{equation}
where $\Psi$, $\tau$, $t$ are the envelope of the waveform, fast time and slow time respectively; $sn$, $cn$, $dn$ are the Jacobi elliptic functions; $\epsilon$ is a free parameter of the solution. $\epsilon$ affects the amplitude of the waveform and the breathing period, but does not affect the waveform shape significantly. Here, we choose $\epsilon$=0.5 as an example of this solution. The corresponding temporal and spectral breathing can be found in Fig. \ref{Fig5DSP}. The spectral dynamics clearly shows the breathing of the solution. However, the features of the``double period" solution differ significantly from the dark breather we have investigated in microresonators. This breathing solution does not resemble the dark-pulse-like waveform in the microresonator (Fig. \ref{Fig5DSP}c). In the frequency domain, there is also energy exchange between the 0-frequency and the frequency components in the wings ($\pi$ phase difference). However, all the frequency components in the wing (i.e., aside from the 0-frequency) breathe with the same phase (Fig. \ref{Fig5DSP}d), which differs from the breathing of dark breather presented in Fig. 2c of the main text.

The NLSE is usually used to describe conservative systems, such as the propagation in optical fibers. The governing equation of the Kerr comb generation in microresonators is the Lugiato-Lefever equation (LLE) \cite{Lugiato_PRL1987spatial,Coen_OL2013modeling}, which is also regarded as an externally driven damped NLSE \cite{Smirnov_PRE1996}. The distinct breather dynamics in the NLSE and the LLE in the normal dispersion regime highlight the importance of dissipation for Kerr comb generation in microresonators.

\section*{5. Dark breather at the absence of mode-interaction}

\begin{figure}[h]
\includegraphics[width=0.75\columnwidth]{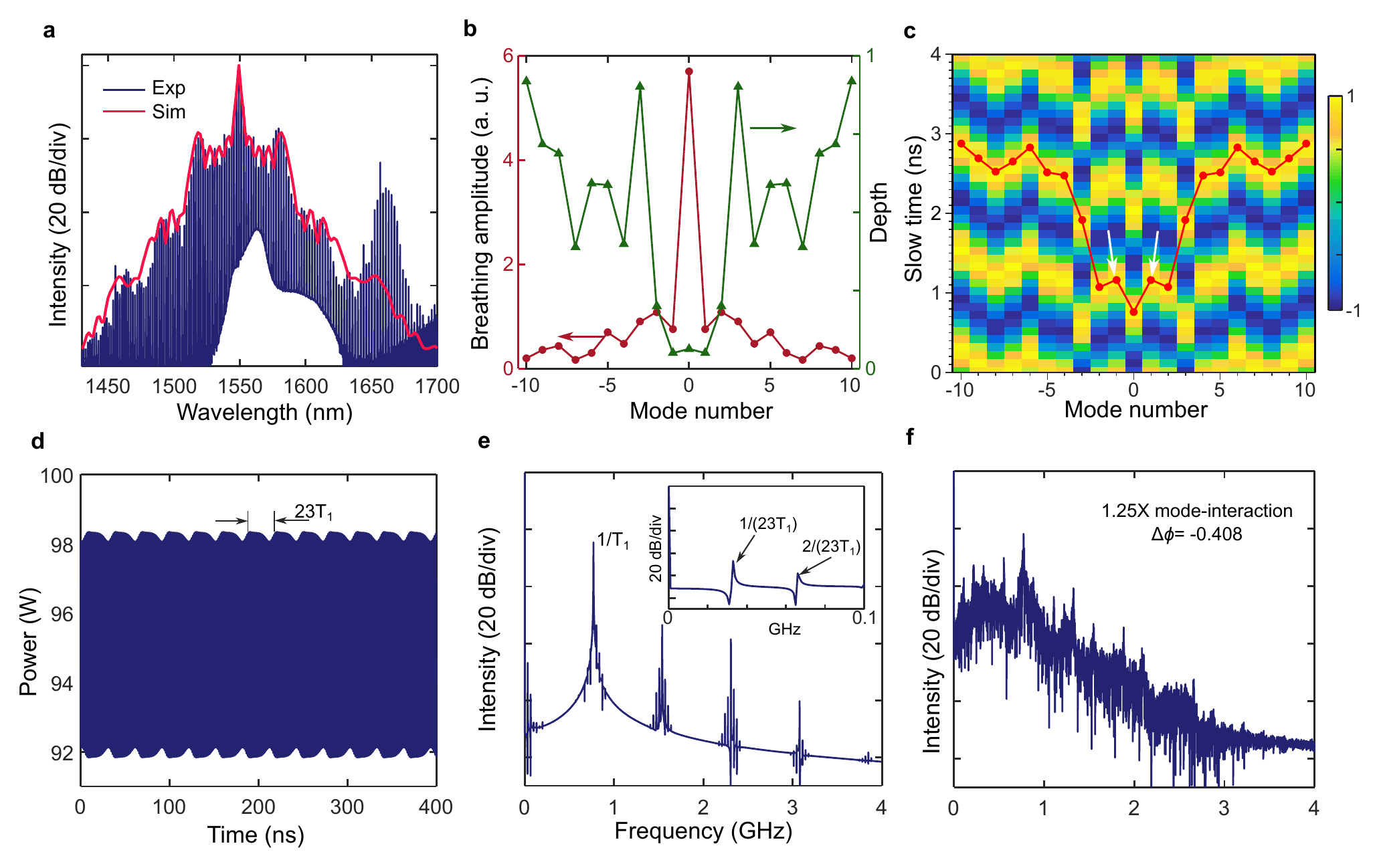}
\caption{\textbf{Dark breather at the absence of mode-interaction.} \textbf{a,} Simulated (averaged over slow time) spectrum in comparison with the experimental measured spectrum of the dark breather. \textbf{b,} Absolute breathing amplitude and breathing depth of different comb lines, both of which are symmetric with respect to the pump.  \textbf{c,} Spectral breathing dynamics for different comb lines, which are also symmetric with respect to the pump (see the arrows pointed to modes $\pm$1). The circles on the red line mark the peaks of the spectral breathing. \textbf{d,} In the absence of mode-interaction, the comb simulated at 752 mW pump power and $\delta_0$=0.0629 shows an oscillation period 23T$_1$ (where T$_1$ is the ``period" for a period-1 breather). This is in contrast to the period-3 bifurcation simulated in the presence of the mode interaction. \textbf{e,} Corresponding RF spectrum of \textbf{d}; the inset reveals low frequency RF tones at 1/(23T$_1$) and 2/(23T$_1$). \textbf{f,} If the mode-interaction strength is increased to 1.25 times the value used in the main text (i.e., $\Delta\phi=-$0.408), the simulated RF spectrum becomes broadband (using 752 mW pump power and $\delta_0$=0.0629, same as parts \textbf{d-e}).}
\label{Fig4NoMI}
\end{figure}

The simulated spectral breathing becomes symmetric with respect to the pump, when the mode-interaction is excluded (i.e., setting $\Delta \phi$=0). Note that this simulation uses the dark breather described in Fig. 4a of the main text as the seed. The simulated spectrum of the dark breather (averaged over slow time) no longer shows spikes around 1657 nm (Fig. \ref{Fig4NoMI}a). The breathing amplitude of different comb lines becomes symmetric with respect to the pump (Fig. \ref{Fig4NoMI}b). Furthermore, the breathing phase of different lines becomes symmetric with respect to the pump. For instance, modes $\pm$1 (see the arrows in (Fig. \ref{Fig4NoMI}c)) breathe with the same phase, whereas modes $\pm$1 breathe with different phase in the presence of mode-interaction (see Fig. 2c and Fig. 4g in the main text).

Furthermore, mode-interaction strongly affects the occurrence of the period-3 bifurcation reported in the main text. For instance, in the absence of mode-interaction, at the pump power of 752 mW, $\delta_0$=0.0629 where the simulated dark breather exhibits the period-3 bifurcation with mode-interaction in the main text, the simulated dark breather no longer shows the period-3 bifurcation (Figs. \ref{Fig4NoMI}d, e). Moreover, no period-3 bifurcation is found in simulation even at other pump powers in the absence of mode-interaction. In addition, if we increase the mode-interaction strength to 1.25 times that used in the main text, the dark breather will now be in the chaotic breathing state at 752 mW pump power, $\delta_0$=0.0629. We repeat simulations at other pump powers but do not find a period-3 bifurcation in the presence of this increased mode-interaction strength either. These results suggest that mode-interactions strongly affect the period bifurcation of the dark breather and are critical to the observation of the period-3 bifurcation reported in the main text.

\section*{6. Illustration of animations}
\textbf{Animation 1,} spectral and temporal breathing of the regular breather (corresponds to Fig. 4a in the main text). The breather repeats itself around 1.2$\sim$1.3 ns and $\sim$2.5 ns (end of the animation), as the period is 1.27 ns.

\textbf{Animation 2,} spectral and temporal breathing of the chaotic breather (corresponds to Fig. 5e in the main text). The chaotic breather does not repeat itself in the full 2.5 ns.

\bibliography{reflist}
\parskip 12pt